\documentclass[reprint,aps,prb]{revtex4-2}

\usepackage[pdftex]{graphicx}
\usepackage{amsmath}
\usepackage{tabularx}
\usepackage{colortbl}
\usepackage{etoolbox}
\usepackage[dvipsnames]{xcolor}
\usepackage{enumitem}
\usepackage{array}
\usepackage{makecell}

\makeatletter
\newlength{\qrr@dimen@}
\expandafter\pretocmd\csname tabular*\endcsname{\setlength{\qrr@dimen@}{#1}}{}{}
\newcommand*{\Rowcolor}[2][\tabcolsep]{%
	\ifx\relax#1\relax\else
	\kern-\the\dimexpr#1\relax
	\fi
	\makebox[0pt][l]{%
		\fboxsep=0pt
		\colorbox{#2}{%
			\strut\kern\qrr@dimen@
		}%
	}%
}

\definecolor{mygray}{gray}{0.9}

\begin{document}

\title{Combinatorial exploration of quantum spin liquid candidates in the herbertsmithite material family}

\author{Alex Hallett}
\affiliation{Materials Department, University of California, Santa Barbara, California 93106, USA}

\author{Catalina Avarvarei}
\affiliation{Materials Department, University of California, Santa Barbara, California 93106, USA}

\author{John W. Harter}
\email[Corresponding author: ]{harter@ucsb.edu}
\affiliation{Materials Department, University of California, Santa Barbara, California 93106, USA}

\date{\today}

\begin{abstract}
Geometric frustration of magnetic ions can lead to a quantum spin liquid ground state where long range magnetic order is avoided despite strong exchange interactions. The physical realization of quantum spin liquids comprises a major unresolved area of contemporary materials science. One prominent magnetically-frustrated structure is the kagome lattice. The naturally occurring minerals herbertsmithite [ZnCu$_3$(OH)$_6$Cl$_2$] and Zn-substituted barlowite [ZnCu$_3$(OH)$_6$BrF] both feature perfect kagome layers of spin-$1/2$ copper ions and display experimental signatures consistent with a quantum spin liquid state at low temperatures. To investigate other possible candidates within this material family, we perform a systematic first-principles combinatorial exploration of structurally related compounds [$A$Cu$_3$(OH)$_6B_2$ and $A$Cu$_3$(OH)$_6BC$] by substituting non-magnetic divalent cations ($A$) and halide anions ($B$, $C$). After optimizing such structures using density functional theory, we compare various structural and thermodynamic parameters to determine which compounds are most likely to favor a quantum spin liquid state. Convex hull calculations using binary compounds are performed to determine feasibility of synthesis. We also estimate the likelihood of interlayer substitutional disorder and spontaneous distortions of the kagome layers. After considering all of these factors as a whole, we select several promising candidate materials that we believe deserve further attention.
\end{abstract}

\maketitle

\section{Introduction}

\begin{figure}[ht]
\includegraphics[width=\columnwidth]{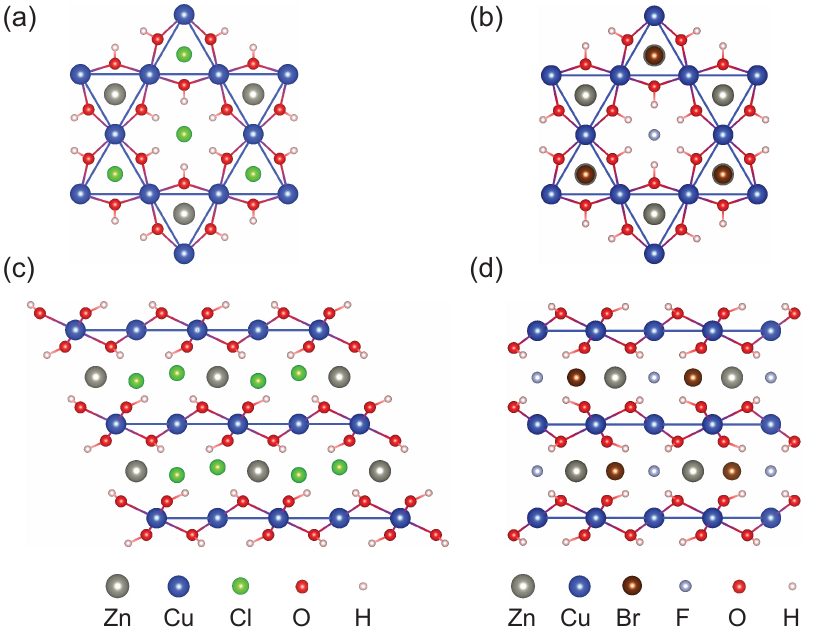}
\caption{Crystal structures of herbertsmithite and Zn-barlowite. (a)~Herbertsmithite viewed along the $c$-axis, showing the kagome arrangement of Cu ions. (b)~Zn-barlowite viewed along the $c$-axis. (c)~Herbertsmithite viewed along the [110] direction, showing the shifted stacking arrangement of the kagome layers. (d)~Zn-barlowite viewed along [110], showing the stacking of the kagome layers and the inequivalence of the Br and F sites.}
\label{fig:Fig_1}
\end{figure}

In a quantum spin liquid (QSL), frustrated antiferromagnetic exchange interactions prevent localized spins from ordering at low temperatures, instead forming a fluid-like phase. The large degeneracy of this state can give rise to novel phenomena such as fractionalized quasiparticles, emergent gauge fields, and long-range entanglement~\cite{balentsSpinLiquidsFrustrated2010,savaryQuantumSpinLiquids2016,broholmQuantumSpinLiquids2020,semeghiniProbingTopologicalSpin2021}. The kagome lattice of corner-sharing triangles is known to have high geometric frustration and is capable of hosting such a phase. A leading QSL material candidate possessing this structure is herbertsmithite [ZnCu$_3$(OH)$_6$Cl$_2$], which contains perfect kagome layers of spin-$1/2$ copper cations separated by non-magnetic Zn and Cl ions~\cite{mendelsQuantumKagomeAntiferromagnet2010,normanColloquiumHerbertsmithiteSearch2016a}, as shown in Fig.~\ref{fig:Fig_1}(a,c). Indeed, although herbertsmithite has strong antiferromagnetic exchange interactions, no magnetic phase transition is observed down to sub-kelvin temperatures~\cite{shoresStructurallyPerfectKagome2005,devriesScaleFreeAntiferromagneticFluctuations2009,hanSynthesisCharacterizationSingle2011,heltonSpinDynamicsSpin2007,mendelsQuantumMagnetismParatacamite2007}, and an array of experimental and theoretical work favors a possible QSL scenario~\cite{fuEvidenceGappedSpinliquid2015,hanCorrelatedImpuritiesIntrinsic2016,olariu17MathrmONMR2008,hanFractionalizedExcitationsSpinliquid2012,wulferdingInterplayThermalQuantum2010,devriesMagneticGroundState2008,khuntiaGaplessGroundState2020,oferHerbertsmithiteHamiltonianMSR2011,jeschkeFirstprinciplesDeterminationHeisenberg2013,suttnerRenormalizationGroupAnalysis2014,jansonDFTbasedMicroscopicMagnetic, rigolMagneticSusceptibilityKagome2007, oferHerbertsmithiteHamiltonianMSR2011,gotzeRouteMagneticOrder2016}. 

Despite its many promising features, herbertsmithite is prone to cation substitutional disorder, where Cu may occupy interlayer sites and Zn may occupy intralayer kagome sites~\cite{shoresStructurallyPerfectKagome2005,olariu17MathrmONMR2008,liLatticeDynamicsSpin2020}. The precise amount of this disorder is debated. Several studies suggest that while there is minimal substitution of Zn on the kagome layers, the interlayer sites can be occupied by up to 15\% Cu~\cite{freedmanSiteSpecificXray2010,imaiLocalSpinSusceptibility2011,fuEvidenceGappedSpinliquid2015,devriesExtensionZincParatacamite2012}, resulting in a decidedly off-stoichiometric compound. These interlayer ``orphan'' spin-$1/2$ Cu$^{2+}$ defects are highly problematic for the QSL state, causing weak ferromagnetic interactions between kagome layers and distorting the surrounding matrix of magnetic ions~\cite{hanCorrelatedImpuritiesIntrinsic2016}. Zn-substituted barlowite (Zn-barlowite), a structurally related compound and another potential QSL candidate~\cite{fengGappedSpin1Spinon2017,tustainMagneticOrderQuantum2020}, is thought to have a much lower interlayer disorder concentration, largely due to the greater chemical distinction between the interlayer and intralayer sites, as shown in Fig.~\ref{fig:Fig_1}(b,d)~\cite{liuSelectivelyDopingBarlowite2015,guterdingReductionMagneticInterlayer2016}. Experiments indicate that in Zn-barlowite, off-center interlayer $C_{2v}$ sites can contain up to ~5\% Cu defects. Like herbertsmithite, however, Zn-barlowite does not order magnetically, even with these large concentrations of magnetic defects~\cite{smahaSiteSpecificStructureMultiple2020,smahaMaterializingRivalGround2020,pascoBarlowiteNextQuantum2018}. While progress on this class of materials is encouraging, it is nevertheless desirable to further minimize orphan Cu spins to realize a clean QSL ground state.

Synthesizing compounds structurally similar to herbertsmithite and Zn-barlowite is a promising route to discover new QSL candidates. For example, Mg-substituted herbertsmithite, Mg$_x$Cu$_{4-x}$(OH)$_6$Cl$_2$ (tondiite), has been successfully synthesized and shows no magnetic phase transition down to 1.8~K~\cite{chuCu2KagomeAntiferromagnet2010,colmanComparisonsHaydeeiteACu3Mg2010,malcherekTondiiteCu3MgOH2014}, and a Cd analog [CdCu$_3$(OH)$_6$Cl$_2$] shows no magnetic ordering down to 2~K, although it exhibits significant distortions of the kagome planes~\cite{mcqueenCdCu3OH6Cl22011}. Synthesis of the bromide analog of herbertsmithite [ZnCu$_3$(OH)$_6$Br$_2$] was attempted but unsuccessful~\cite{braithwaiteHerbertsmithiteCu3ZnOH2004}. A Zn-barlowite related structure, Zn-claringbullite [ZnCu$_3$(OH)$_6$ClF], shows no obvious magnetic transition down to 2~K, but a perfectly stoichiometric compound was not achieved~\cite{fengClaringbulliteNewSpin2018}. While the Mg analog of barlowite cannot be synthesized due to the insolubility of MgF$_2$ in water, the bromide analog was attempted [MgCu$_3$(OH)$_6$Br$_2$], but did not have the Zn-barlowite structure and ordered antiferromagnetically at 5.4~K~\cite{weiAntiferromagnetismKagomelatticeCompound2019}.

Clearly, more work is needed to search for and identify viable candidates in this material family. Only a few computational studies exist exploring cation substitution in barlowite~\cite{liuSelectivelyDopingBarlowite2015,guterdingReductionMagneticInterlayer2016}, and a complete exploration of the structural families of herbertsmithite and Zn-barlowite using computational methods has not been performed. In this paper, we use \textit{ab initio} calculations to systematically explore compounds within the herbertsmithite and Zn-barlowite families. We compare the thermodynamic stability, structural properties, and tendency towards disorder. After considering all these criteria together, we select promising QSL candidates that merit further experimental and theoretical examination. 

\section{Computational Procedure}

We carry out a systematic exploration of the structural relatives of herbertsmithite [$A$Cu$_3$(OH)$_6B_2$] and Zn-barlowite [$A$Cu$_3$(OH)$_6BC$] by substituting closed-shell (spinless) $2+$ cations ($A$ = Ba, Be, Ca, Cd, Ge, Hg, Mg, Pb, Sn, Sr, Zn) and halide anions ($B,C$ = Br, Cl, F, I). We investigate all 44 possible herbertsmithite relatives. While there are 176 possible Zn-barlowite relatives, we eliminate compounds where $B = C$ because the herbertsmithite structure always has lower energy in these cases. We also do not consider compounds in which the less electronegative anion occupies the $C$ site [the site occupied by F in  Fig.~\ref{fig:Fig_1}(b,d)]. All hydrogen bonds are oriented towards the $C$ site, so the more electronegative ion will always occupy this position to minimize energy. Thus, a total of 66 relatives in the Zn-barlowite family were selected for consideration. 

We perform high-throughput calculations where the structural optimization of each candidate is followed by a static calculation to extract the ground-state energy and to compute phonon frequencies at the $\Gamma$ point to confirm structural stability. In addition to confirming the stability of the relaxed structures, we perform convex hull calculations to determine if synthesis of the candidate compounds is thermodynamically feasible. For the most promising materials, we also calculate defect formation energies and full phonon dispersions throughout the first Brillouin zone to verify stability at $k$-points away from the zone center.  

All structures were calculated by allowing the lattice parameters, cell volume, and atomic positions to fully relax using density functional theory (DFT) as implemented in the Vienna \textit{ab initio} simulation package (\textsc{vasp})~\cite{kresseInitioMolecularDynamics1993, kresseEfficientIterativeSchemes1996, kresseEfficiencyAbinitioTotal1996}. We used the supplied projector augmented wave potentials~\cite{kresseUltrasoftPseudopotentialsProjector1999} within the generalized gradient approximation and Perdew-Burke-Ernzerhof scheme~\cite{perdewGeneralizedGradientApproximation1996}. Electronic wave functions were expanded in a plane wave basis set with an energy cutoff of 800~eV, and reciprocal space was sampled using an $8\times8\times8$ $k$-point mesh for herbertsmithite-related structures and an $8\times8\times5$ $k$-point mesh for Zn-barlowite-related structures. A $\Gamma$-centered mesh is necessary due to the hexagonal symmetry of Zn-barlowite. The spacing between $k$-points was $\sim$0.15~\AA$^{-1}$ for both structural families, and this spacing was also used for calculating the energies of binary compounds used in the convex hull analysis. All structures were relaxed until forces on the atoms were less than 1~meV/\AA. Calculations were non-spin-polarized. Input files for all calculations can be found in the Supplemental Material~\cite{SeeSupplementalMaterial}.

\begin{figure*}[ht]
\includegraphics[scale=1]{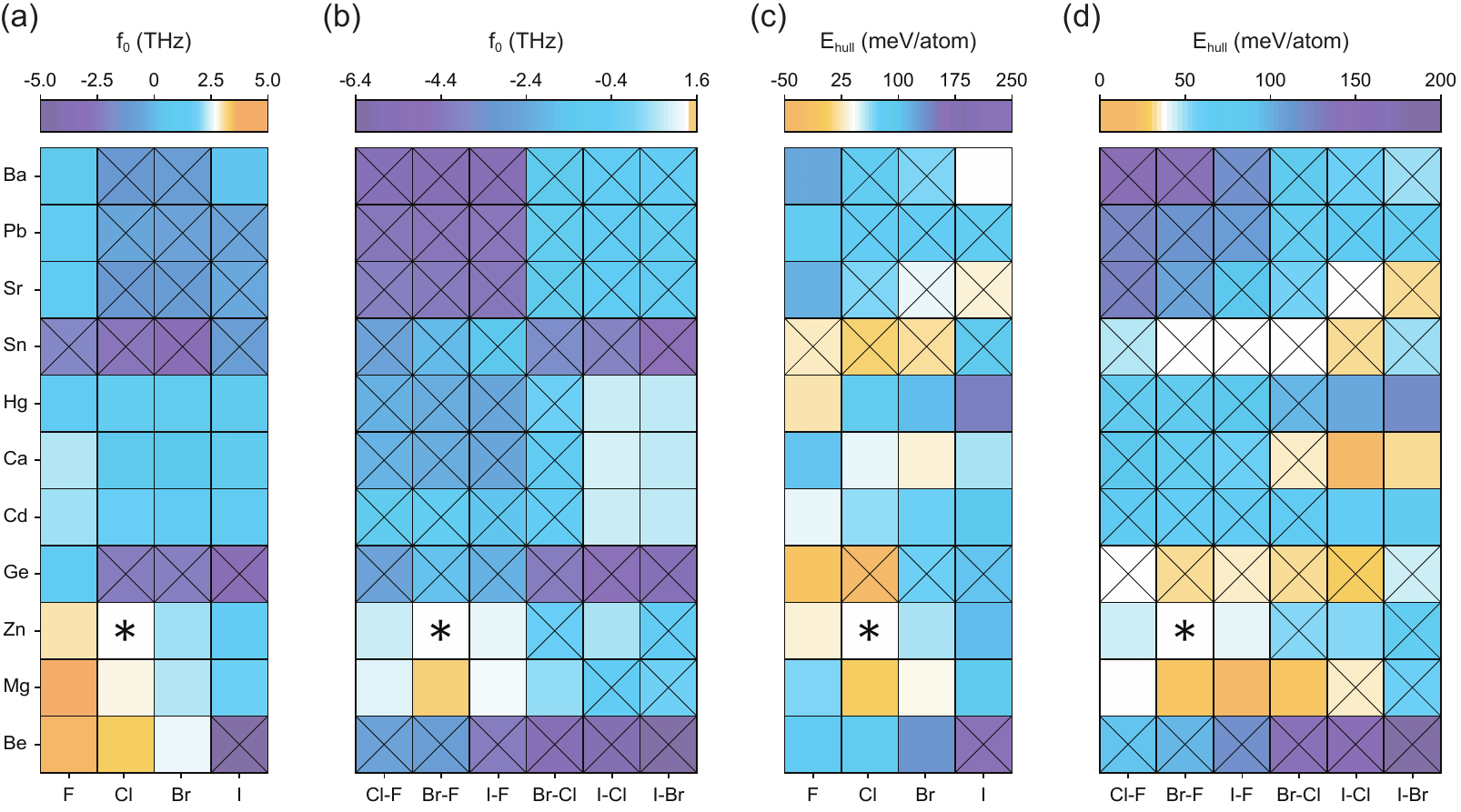}
\caption{Structural stability and thermodynamics of candidate compounds. (a)~Lowest optical phonon frequency for herbertsmithite-related candidates. (b)~Lowest optical phonon frequency for Zn-barlowite-related candidates. (c)~Convex hull energies for herbertsmithite-related candidates. (d)~Convex hull energies for Zn-barlowite-related candidates. Structurally unstable compounds (identified by $f_0 < 0$) are denoted with an `X'. Cations are shown on the vertical axis and anions on the horizontal axis, in order of increasing ionic radius from bottom to top and left to right, respectively. The reference compound (either herbertsmithite or Zn-barlowite) is shown in white and marked with an asterisk. Compounds with parameter values more favorable than the reference compounds are shown with warm colors, and values less favorable are shown with cool colors.}
\label{fig:Fig_2}
\end{figure*} 

\begin{figure*}[ht]
\includegraphics[scale=1]{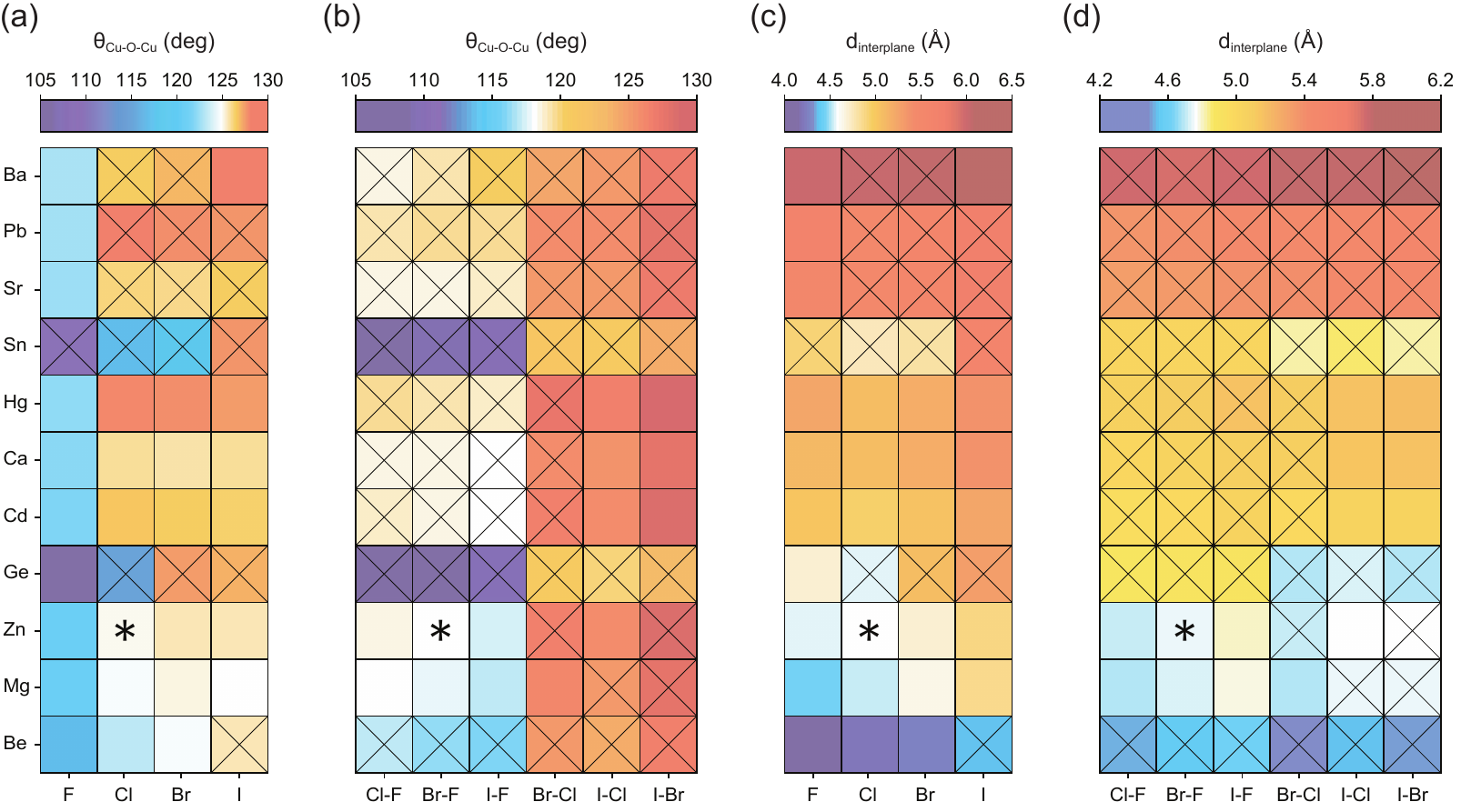}
\caption{Structural properties of candidate compounds. (a)~Cu-O-Cu bond angle for herbertsmithite-related candidates. (b)~Cu-O-Cu bond angle for Zn-barlowite-related candidates. (c)~Interplane kagome distance for herbertsmithite-related candidates. (d)~Interplane kagome distance for Zn-barlowite-related candidates. Structurally unstable compounds are denoted with an `X'. Cations are shown on the vertical axis and anions on the horizontal axis, in order of increasing ionic radius from bottom to top and left to right, respectively. The reference compound (either herbertsmithite or Zn-barlowite) is shown in white and marked with an asterisk. Compounds with parameter values more favorable than the reference compounds are shown with warm colors, and values less favorable are shown with cool colors.}
\label{fig:Fig_3}
\end{figure*} 

\begin{figure*}[ht]
\includegraphics[width=\textwidth]{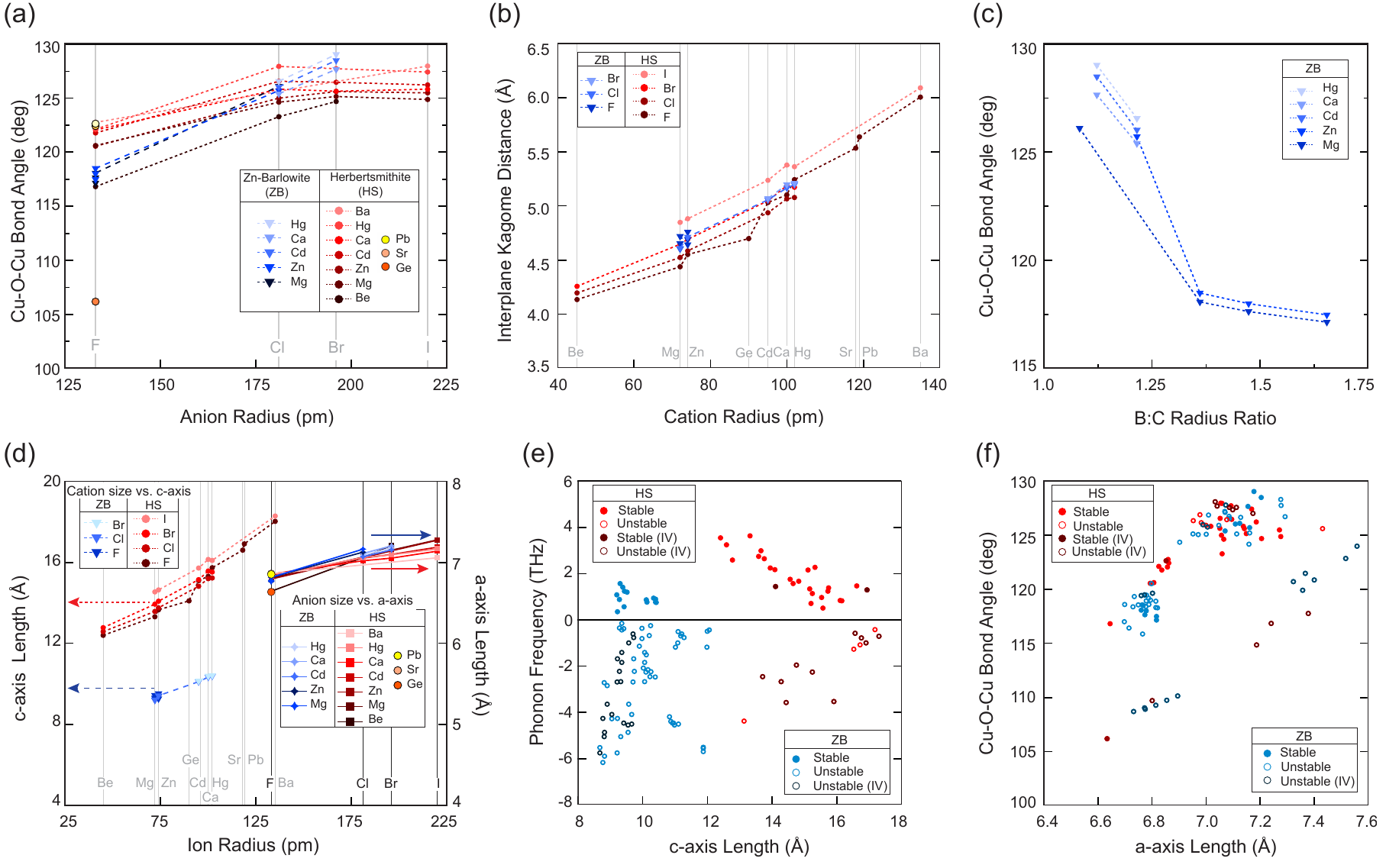}
\caption{Dependence of structural properties on ion size. (a)~Cu-O-Cu bond angle versus anion radius. For Zn-barlowite, the radius plotted is that of the most electronegative anion. Blue (red) traces correspond to Zn-barlowite (herbertsmithite) relatives. Different cations are plotted as separate traces where darker (lighter) traces correspond to smaller (larger) ion sizes. (b)~Interplane kagome distance versus cation radius for herbertsmithite (red) and Zn-barlowite (blue) relatives. Separate traces are plotted for each anion, where small (large) anions are plotted in dark (light) shades. (c)~Cu-O-Cu bond angle versus the anion $B$ to anion $C$ ratio for stable compounds. Separate traces are plotted for different cations. (d)~$c$-axis length versus cation size (left, dashed line) and $a$-axis length versus anion size (right, solid lines). (e)~Frequency of the lowest optical phonon mode versus $c$-axis length for Zn-barlowite (blue) and herbertsmithite (red) relatives. Stable (unstable) compounds are shown with filled (empty) markers. The group IV elements (Ge, Sn, Pb) are plotted with darker colors because they are almost always unstable, regardless of their $c$-axis length. (f)~Cu-O-Cu bond angle versus $a$-axis length for Zn-barlowite (blue) and herbertsmithite (red) relatives. Stable (unstable) compounds are shown with filled (empty) markers. Compounds containing group IV cations (shown in darker colors) tend to be unstable and have much smaller bond angles.}
\label{fig:Fig_4}
\end{figure*}

\section{Results and Discussion}

\subsection{Phonon Calculations}

Phonon calculations at the $\Gamma$ point for the fully-relaxed structures were performed in \textsc{vasp} within the finite differences approximation to confirm structural stability. As expected, many structures have unstable phonon modes. Fig.~\ref{fig:Fig_2}(a,b) shows the frequency of the lowest energy optical phonon mode, $f_0$, for all compounds. In all subsequent plots, the unstable compounds (with $f_0 < 0$) are marked with an `X' to distinguish them from structurally stable and potentially viable candidates. Cations are shown on the vertical axis and anions on the horizontal axis, in order of increasing ionic radius from bottom to top and left to right, respectively. The reference compound, either herbertsmithite or Zn-barlowite, is shown in white and marked with an asterisk. Compounds with parameter values more favorable than the reference compound are shown with warm colors, and values less favorable are shown with cool colors. For example, a higher frequency of the lowest energy optical mode indicates higher dynamical stability, so higher frequencies are shown with warm colors. Most compounds containing group IV elements (Ge, Sn, Pb) tend to be unstable, with the exception of GeCu$_3$(OH)$_6$F$_2$ and PbCu$_3$(OH)$_6$F$_2$. Compounds containing larger cations are generally unstable, as well as Zn-barlowite relatives containing Be. 

\subsection{Convex Hull Calculations}

The convex hull of a compound is useful for determining if synthesis is thermodynamically feasible, usually through a comparison of the compound's formation energy to the sum of the energies of all other possible combinations of crystal structures that could be created from the same set of elements in the same ratios. Due to the prohibitive size of the phase space for our candidate materials, we perform a simplified procedure. Instead of considering all possible crystal structures, we consider only simple binary ionic compounds [e.g. $A$(OH)$_2$, $AB_2$], which are most likely to yield the lowest convex hull energies (see Supplemental Material~\cite{SeeSupplementalMaterial}). Starting structures for these binary compounds were obtained from the Materials Project~\cite{jainCommentaryMaterialsProject2013} and then re-relaxed with our settings.

Insulators with energies less than $\sim$50~meV above the convex hull tend to be stable~\cite{jiangScreeningPromisingCsV2022}. We therefore use an energy cutoff of 50~meV/atom as our criteria for thermodynamic stability when identifying candidate materials. The calculated energy above the hull for each compound is shown in Fig.~\ref{fig:Fig_2}(c,d). Energies higher than the reference compound are considered unfavorable and are represented with cool colors, while energies lower than the reference compound are favorable and represented with warm colors. Again, the reference compounds are shown in white and marked with an asterisk, and compounds with structural instabilities (as determined by phonon calculations) are marked with an `X'. There does not appear to be a clear connection between convex hull energy and structural stability or ion size. 

\subsection{Comparing Structural Parameters}

In addition to structural and thermodynamic stability, we use Cu-O-Cu bond angles and spacings between kagome layers as additional metrics to rank the candidate compounds. A Cu-O-Cu bond angle approaching 180$^{\circ}$ leads to a large antiferromagnetic superexchange interaction while minimizing undesirable Dzyaloshinskii–Moriya interactions. Larger bond angles are therefore highly desirable. Additionally, a greater separation between the kagome layers isolates the two-dimensional magnetic subsystems and suppresses unwanted coupling between planes. In Fig.~\ref{fig:Fig_3}, these two structural properties are displayed for all candidate compounds. Squares corresponding to specific compounds are colored and marked according to the same system described for Fig.~\ref{fig:Fig_2}, where bond angles and interplane distances larger (smaller) than the reference compounds are favorable (unfavorable) and represented with warm (cool) colors, and structurally unstable compounds continue to be marked with an `X'. Compounds with larger cation and anion radii generally lead to larger bond angles and interplane distances, but also tend to be structurally unstable. Compounds containing group IV elements are unstable and tend to have smaller bond angles.

In Fig.~\ref{fig:Fig_4}, we investigate the effects of ion size on the physical properties of the candidate compounds in more detail. In Fig.~\ref{fig:Fig_4}(a), the Cu-O-Cu bond angle is plotted versus anion radius for the structurally stable materials. The anion size plotted on the horizontal axis for Zn-barlowite relatives refers to the $C$-site anion that occupies the same position as F in the reference compound [ZnCu$_{3}$(OH)$_{6}$BrF] because it has the largest influence on bond angle. For all materials, bond angle increases with increasing anion size, and for a given anion, the bond angle also increases with increasing cation size. Figure~\ref{fig:Fig_4}(b) shows the kagome plane spacing versus cation radius for stable compounds, with separate traces for each anion. As expected, a larger cation radius leads to greater distance between the kagome layers. For a given cation, interplane distance also increases with increasing anion size. In Fig.~\ref{fig:Fig_4}(c), we find that while the $C$-site anion has the greatest effect on the Cu-O-Cu bond angle, larger bond angles are obtained when the $B$-site anion is similar in size to the $C$-site anion.

We examine the effect of ion size on the lattice parameters of stable compounds in Fig.~\ref{fig:Fig_4}(d). The $c$-axis length primarily increases with cation size while the $a$-axis length primarily increases with anion size, although anion size has a much weaker affect on the $a$-axis than cation size does on the $c$-axis. The frequency of the lowest optical phonon mode ($f_0$) is plotted against $c$-axis length in Fig.~\ref{fig:Fig_4}(e) for both stable (filled markers) and unstable (empty markers) structures. Of all the structural parameters, the $c$-axis length has the highest correlation with $f_0$. For herbertsmithite relatives, as the $c$-axis increases, $f_0$ decreases, meaning compounds tend to be less dynamically stable. Compounds containing group IV ions (Ge, Sn, Pb) are plotted in darker shades for both structural families because nearly all compounds containing these elements are unstable. Of the compounds not containing group IV ions, $c$-axis lengths that are very small or very large lead to structural instabilities. Compounds containing cations from groups IIA and IIB which are close in size to Zn tend to be most stable. Fig.~\ref{fig:Fig_4}(f) shows Cu-O-Cu bond angle versus $a$-axis length. We find that a larger $a$-axis leads to a larger bond angle, which agrees with the results in Fig.~\ref{fig:Fig_4}(a), where bond angle is positively correlated with anion radius, and Fig.~\ref{fig:Fig_4}(d), which shows the positive correlation between anion size and the length of the $a$-axis. It should be noted that many unstable compounds containing group IV elements have much smaller bond angles than most other candidates. 

We also explored correlations between Cu-O-Cu bond angle, interplane distance, and in-plane Cu-Cu bond length. These plots can be found in the Supplemental Material~\cite{SeeSupplementalMaterial}. The Cu-O-Cu bond angle has a weak positive correlation with interplane distance. There is also a positive correlation between in-plane Cu-Cu distance and Cu-O-Cu bond angle, as both are influenced by the length of the $a$-axis, which increases with increasing anion size. There is no obvious correlation between the interplane kagome distance and the in-plane Cu-Cu bond length, as the interplane distance depends mostly on cation size, and in-plane bond length depends on anion size. Overall, for both structural families, compounds with cations of intermediate size (Mg, Zn, Cd, and Hg) are most stable. Compounds containing group IV elements (Ge, Sn, Pb) are mostly unstable. Larger anions and cations lead to favorable structural properties, such as larger bond angles and interplane distances, but may also lead to distortions of the kagome layers or other structural instabilities. 

\begin{table*}
	\centering
	\caption{Properties of the most promising QSL candidate materials as compared to the reference materials. The references (herbertsmithite and Zn-barlowite) are highlighted in gray, and the final candidates (with no instabilities throughout the Brillouin zone) are marked with asterisks.}
	\begin{tabularx}{\textwidth}{l@{\extracolsep{\fill}} c c c c c c}
		\hline
		\hline
		Compound  & $f_{0}$ (THz) & $E_\mathrm{hull}$ (meV/atom) & $E_d^f$ (eV) & $\theta$ (deg) & $d_\mathrm{inter}$ (\AA) & $d_\mathrm{in}$ (\AA) \\
		\hline
		BaCu$_3$(OH)$_6$I$_2$													& 0.41 & 42.6 & 2.42 & 128.0 & 6.09 & 3.53  \\
		CaCu$_3$(OH)$_6$Br$_2$												& 0.50 & 30.7 & 0.87 & 125.7 & 5.19 & 3.53  \\   
		CaCu$_3$(OH)$_6$Cl$_2$												& 0.70 & 44.8 & 0.57 & 125.8 & 5.06 & 3.51  \\
		MgCu$_3$(OH)$_6$Br$_2$ \textasteriskcentered	& 2.23 & 36.0 & 0.36 & 125.2 & 4.65 & 3.57  \\ 
		\Rowcolor{mygray}
		ZnCu$_3$(OH)$_6$Cl$_2$												& 2.63 & 41.2 & 0.13 & 125.0 & 4.58 & 3.53  \\
		\hline
		CaCu$_3$(OH)$_6$IBr														& 0.77 & 31.6 & 0.74 & 127.7 & 5.20 & 3.58  \\
		CaCu$_3$(OH)$_6$ICl \textasteriskcentered			& 0.94 & 19.2 & 0.72 & 125.4 & 5.17 & 3.54  \\
		MgCu$_3$(OH)$_6$ClF \textasteriskcentered			& 1.09 & 39.6 & 0.39 & 118.1 & 4.60 & 3.38  \\
		MgCu$_3$(OH)$_6$BrCl													& 0.35 & 26.9 & 0.30 & 126.1 & 4.61 & 3.56  \\
		\Rowcolor{mygray}
		ZnCu$_3$(OH)$_6$BrF														& 1.41 & 38.6 & 0.10 & 118.0 & 4.69 & 3.39  \\
		ZnCu$_3$(OH)$_6$ClF														& 0.89 & 43.1 & 0.07 & 118.5 & 4.64 & 3.38  \\
		\hline
		\hline
	\end{tabularx}
\label{table1}
\end{table*} 

\subsection{Defect Formation Energy}

Herbertsmithite and Zn-barlowite are both susceptible to cation disorder. In herbertsmithite, the Jahn-Teller active $d^9$ Cu$^{2+}$ ion occupies the tetragonally elongated site in the center of the CuO$_{4}$Cl$_{2}$ octahedra. The $d^{10}$ Zn$^{2+}$ ions are not Jahn-Teller active, and occupy the higher-symmetry trigonally compressed octahedral sites between the kagome layers. Due to the electronic configurations of the ions and distinct coordination environments, it is not favorable for Zn to occupy the in-plane sites within the kagome layer. However, herbertsmithite is the $x=1$ end member of the Zn-paratacamite family [Zn$_{x}$Cu$_{4-x}$(OH)$_{6}$Cl$_{2}$], and there is a preference for some Cu to exist on the interlayer site instead of full occupation with Zn alone~\cite{shoresStructurallyPerfectKagome2005}. The equilibrium occupation of the interlayer site by Cu has been estimated to be as large as 15\% in herbertsmithite~\cite{freedmanSiteSpecificXray2010,imaiLocalSpinSusceptibility2011}. 

In Zn-barlowite, the interlayer site has a trigonal prismatic geometry, making it even less favorable for the Jahn-Teller active Cu$^{2+}$ ion. As a result, the interlayer Cu occupation is only $\sim$5\% in Zn-barlowite~\cite{smahaSiteSpecificStructureMultiple2020}, confirming early computational predictions~\cite{liuSelectivelyDopingBarlowite2015,guterdingReductionMagneticInterlayer2016}. Site-specific x-ray diffraction measurements have shown that there are two distinct interlayer sites in Zn-barlowite: an off-center $C_{2v}$ site and a central $D_{3h}$ site. The interlayer Cu defects occupy the $C_{2v}$ sites. It should be noted that even for large concentrations of magnetic impurities on the interlayer site, Zn-barlowite does not show signs of magnetic ordering, indicating that the possible QSL phase is somewhat robust against interlayer magnetic impurities~\cite{smahaSiteSpecificStructureMultiple2020}. 

An ideal QSL candidate will have only non-magnetic ions on the interlayer sites, and therefore must have a high energy cost for interlayer Cu substitution. We calculated the formation energy of such defects in a select number of our most promising candidates (those structurally stable, with $E_\mathrm{hull} < 50$~meV/atom, and with bond angles and interplane distances larger than the reference compounds). Since nearly all experimental and computational studies indicate that there is negligible substitution of non-magnetic ions within the kagome layers, we consider only interlayer defects. The general expression for the formation energy of a charge-neutral substitutional defect is 
$${E_{d}^{f} = E[\mathrm{defect}] - E[\mathrm{bulk}] + (\mu_A - \mu_\mathrm{Cu}) = \Delta E_s + \Delta\mu,}$$
where $\Delta E_s$ is the difference in energy between a structure with a single defect and the pristine bulk structure and $\Delta\mu$ is the chemical potential difference of $A$ and Cu. To calculate $E[\mathrm{defect}]$, we construct defect structures from $2\times2\times2$ supercells of herbertsmithite relatives and $2\times2\times1$ supercells of Zn-barlowite relatives, with a single Cu substitution. A depiction of our defect configuration can be found in the Supplemental Material~\cite{SeeSupplementalMaterial}. We relax the atomic positions of the defect structures and subtract the energy of the original defect-free structure to obtain $\Delta E_s$. 

\begin{figure*}
\includegraphics{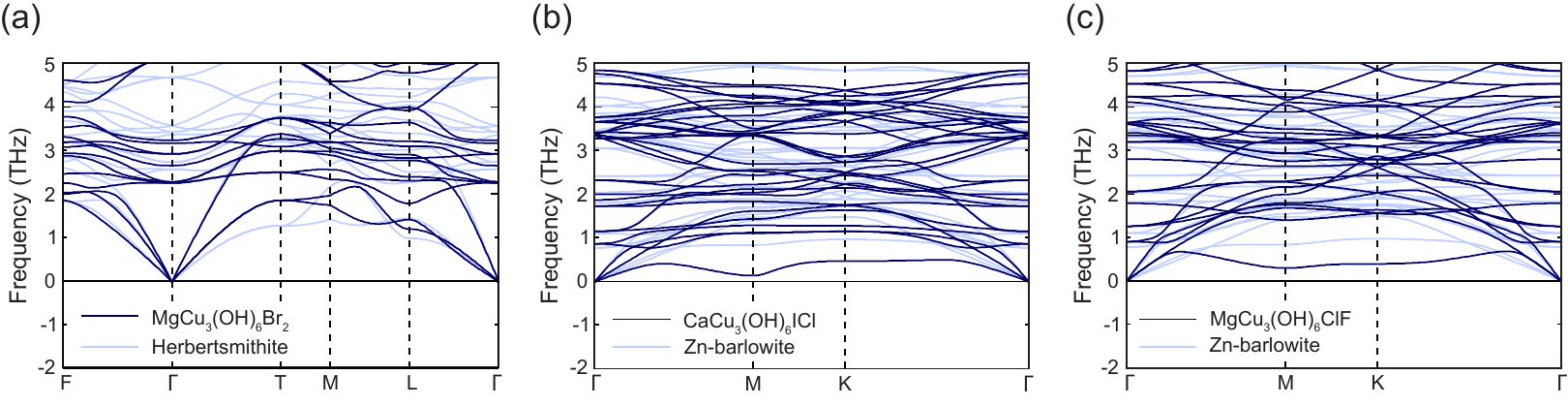}
\caption{Phonon dispersions of final candidates. (a)~The phonon dispersion for MgCu$_3$(OH)$_6$Br$_2$ (blue) overlaid with the reference dispersion for herbertsmithite (gray). (b)~The phonon dispersion for CaCu$_3$(OH)$_6$ICl (blue) overlaid with the reference dispersion for Zn-barlowite (gray). (c)~The phonon dispersion for MgCu$_3$(OH)$_6$ClF (blue) overlaid with the reference dispersion for Zn-barlowite (gray). The absence of imaginary phonon frequencies in all three cases confirms the structural stability of these candidate compounds.}
\label{fig:Fig_5}
\end{figure*}

The chemical formulas for the defect-containing and defect-free configurations are not equivalent, so the chemical potential difference $\Delta\mu = \mu_A - \mu_\mathrm{Cu}$ must be considered. Interlayer defects are primarily created during the initial growth of the material. During synthesis of $A$Cu$_3$(OH)$_6B_2$, the chemical potentials of the constituent elements must satisfy the inequality
$${\mu_A + 3\mu_\mathrm{Cu} + 6\mu_\mathrm{OH} + 2\mu_B > E[A\mathrm{Cu}_3(\mathrm{OH})_6B_2].}$$
Individual chemical potentials must all be less than zero ($\mu_A < 0$, $\mu_B < 0$, $\mu_\mathrm{OH} < 0$, and $\mu_\mathrm{Cu} < 0$). Additionally, the formation of unwanted side products must be avoided, imposing the additional inequalities
$${\mu_A + 2\mu_B < E[AB_2],}$$
$${\mu_\mathrm{Cu} + 2\mu_B < E[\mathrm{Cu}B_2],}$$
$${\mu_A + 2\mu_\mathrm{OH} < E[A(\mathrm{OH})_2].}$$
Similar inequality constraints exist for $A$Cu$_3$(OH)$_6BC$. A higher defect formation energy is preferable to minimize disorder. To maximize $E_{d}^{f}$, we must maximize the chemical potential difference $\Delta\mu$ subject to the above inequality constraints. The defect formation energies calculated with these optimal values of $\Delta\mu$ are given in Table~\ref{table1}. All candidate compounds investigated had a higher energy cost for interlayer defects than herbertsmithite and Zn-barlowite except ZnCu$_3$(OH)$_{6}$ClF (Zn-substituted claringbullite).

Two previous computational studies investigated doping selectivity in barlowite~\cite{liuSelectivelyDopingBarlowite2015,guterdingReductionMagneticInterlayer2016}. In both cases, the authors investigated the likelihood of substituting various non-magnetic ions into the interlayer and intralayer sites of barlowite, in contrast to the present work where we examine the energy cost of a Cu defect on an interlayer site in fully-substituted $A$-barlowite ($A$ = Zn, Mg, Ca). Despite differences in the methodology used to construct defect structures and calculate the chemical potential differences, our findings are generally consistent with those studies, which suggested Zn and Mg to be the most favorable ions for synthesizing barlowite-related compounds. More details on our defect formation energy calculations can be found in the Supplemental Material~\cite{SeeSupplementalMaterial}.

\subsection{Selecting Promising Candidates}

After eliminating all compounds with structural instabilities at the $\Gamma$ point, formation energies greater than 50~meV/atom above the convex hull, and Cu-O-Cu bond angles smaller than the reference compounds, 9 candidate materials remained. For these candidates, we calculated the defect formation energy $E_d^f$. To determine a final ranking, we used the following criteria:
 \begin{enumerate}[noitemsep]
 	\item Structural stability ($f_{0} > 0$)
 	\item Convex hull energy (E$_\mathrm{hull} < 50$~meV/atom)
 	\item Defect energy cost ($E_d^f[\mathrm{candidate}] > E_d^f[\mathrm{ref}]$)
 	\item Cu-O-Cu bond angle ($\theta > \theta^\mathrm{ref}$)
 \end{enumerate}
All compounds satisfying these criteria are listed with their associated properties in Table~\ref{table1}. Complete data sets for all 44 herbertsmithite relatives and 66 Zn-barlowite relatives can be found in the Supplemental Material~\cite{SeeSupplementalMaterial}.
 
We also verified structural stability by calculating the full phonon dispersion throughout the entire Brillouin zone using the finite displacement method within the \textsc{phonopy} code~\cite{togoFirstPrinciplesPhonon2015}. Such calculations can identify structural instabilities associated with an enlargement of the unit cell. Dispersion curves were calculated for all candidates in Table~\ref{table1}. However, only one compound in the herbertsmithite family and two compounds in the Zn-barlowite family were found to be stable throughout the entire Brillouin zone. The dispersion curves of these compounds are shown in Fig.~\ref{fig:Fig_5}, while dispersions for all compounds in Table~\ref{table1} can be found in the Supplemental Material~\cite{SeeSupplementalMaterial}. Surprisingly, while Zn-claringbullite [ZnCu$_3$(OH)$_6$ClF] is known to have perfect kagome layers at room temperature~\cite{fengClaringbulliteNewSpin2018}, our ground state dispersion shows instabilities at the $M$ and $K$ points (see Supplemental Material~\cite{SeeSupplementalMaterial}). The instabilities we observe in DFT may be avoided by thermal fluctuations at room temperature, which could explain the discrepancy between our calculations and the experimental results. Two other Zn-barlowite-related candidate compounds listed in Table~\ref{table1}, CaCu$_3$(OH)$_6$IBr and MgCu$_3$(OH)$_6$BrF, showed similar instabilities, and therefore may also be stable at room temperature (see Supplemental Material~\cite{SeeSupplementalMaterial}).

Our calculations identify MgCu$_3$(OH)$_6$Br$_{2}$ as a potential candidate within the herbertsmithite family, as well as CaCu$_3$(OH)$_6$ICl and MgCu$_3$(OH)$_6$ClF in the Zn-barlowite family. However, some practical considerations related to synthesis may require further investigation. For instance, the Mg analog of Zn-barlowite [MgCu$_3$(OH)$_6$BrF] has not been synthesized due to the insolubility of MgF$_2$ in water. While synthesis of Zn-barlowite using NH$_4$F yields a structurally equivalent compound, crystals obtained using this method show a similar magnetic transition to barlowite, suggesting possible differences in defect structures between the two synthesis methods~\cite{smahaSynthesisdependentPropertiesBarlowite2018}. The insolubility of MgF$_2$ may therefore present difficulty in synthesizing our candidate MgCu$_3$(OH)$_6$ClF~\cite{fengClaringbulliteNewSpin2018}. Synthesis of MgCu$_3$(OH)$_6$Br$_2$ has been attempted, but the desired product was a Zn-barlowite analog~\cite{weiAntiferromagnetismKagomelatticeCompound2019}. The synthesis method, which followed the typical hydrothermal procedure, resulted in a compound with $P\bar3m1$ symmetry, which may mean that the herbertsmithite $R\bar{3}m$ structure is not favored in this reaction. It is possible that other synthesis methods could yield different results. To our knowledge, no experimental studies have been performed on the Ca analog of either herbertsmithite or Zn-barlowite, nor any related compounds containing I. 

\section{Conclusion}

In summary, we performed a systematic combinatorial exploration of herbertsmithite and Zn-barlowite material relatives and identified those with properties that may enhance the likelihood of an ideal QSL ground state. We found several promising candidates---MgCu$_3$(OH)$_6$Br$_2$, CaCu$_3$(OH)$_6$ICl, and MgCu$_3$(OH)$_6$ClF---that are structurally stable, thermodynamically feasible to synthesize, have high energy costs for interlayer defects, and whose structural properties may result in antiferromagnetic superexchange interactions stronger than herbertsmithite or Zn-barlowite. These compounds, if they can be synthesized, may prove to be better QSL candidates than their well-studied counterparts.

\section*{Acknowledgments}

We would like to thank Siavash Karbasizadeh for helpful discussions. This work was supported by the Air Force Office of Scientific Research under AFOSR award no.~FA9550-21-1-0337. C.A. acknowledges support from the UCSB Quantum Foundry Internship Program, which is funded by the National Science Foundation (NSF) through Enabling Quantum Leap: Convergent Accelerated Discovery Foundries for Quantum Materials Science, Engineering, and Information (Q-AMASE-i): Quantum Foundry at UC Santa Barbara (DMR-1906325). Use was made of computational facilities purchased with funds from the NSF (CNS-1725797) and administered by the Center for Scientific Computing (CSC). The CSC is supported by the California NanoSystems Institute and the Research Science and Engineering Center (MRSEC; NSF DMR-1720256) at UC Santa Barbara.


%

\end{document}